# Improving Noise Tolerance of Mixed-Signal Neural Networks

Michael Klachko, Mohammad Reza Mahmoodi, and Dmitri Strukov
Department of Electrical and Computer Engineering
University of California at Santa Barbara
Santa Barbara, USA
{mklachko, mrmahmoodi, strukov}@ucsb.edu

*Abstract—* Mixed-signal hardware accelerators for deep learning achieve orders of magnitude better power efficiency than their digital counterparts. In the ultra-low power consumption regime, limited signal precision inherent to analog computation becomes a challenge. We perform a case study of a 6-layer convolutional neural network running on a mixed-signal accelerator and evaluate its sensitivity to hardware specific noise. We apply various methods to improve noise robustness of the network and demonstrate an effective way to optimize useful signal ranges through adaptive signal clipping. The resulting model is robust enough to achieve 80.2% classification accuracy on CIFAR-10 dataset with just 1.4 mW power budget, while 6 mW budget allows us to achieve 87.1% accuracy, which is within 1% of the software baseline. For comparison, the unoptimized version of the same model achieves only 67.7% accuracy at 1.4 mW and 78.6% at 6 mW.

*Keywords — Mixed-Signal Neural Networks, Vector-Matrix Multipliers, Floating-Gate Memory, Deep Learning, Noise Tolerance*

## I. Introduction

Convolutional neural networks have enjoyed dramatic success recently as the main engine behind deep learning revolution [1-3]. While much effort has been dedicated to developing new, increasingly larger, and more complex network models, the fundamental operation (dot product of two vectors) performed by these models remains unchanged. This operation is extremely resource intensive, consuming up to 95% of the total computational budget of general purpose digital processors (CPU or GPU) [4], with memory bandwidth frequently becoming performance bottleneck [5-6]. The combination of compute and memory transfer bottlenecks makes specialized hardware utilizing in memory computing an attractive platform to accelerate neural network computation.

Recent advances in analog-grade dense nonvolatile memories have enabled extremely fast, compact, and energy efficient analog and mixed-signal circuits [7-9]. Such circuits are perfectly suited for low-to-medium precision dot-product operations. Even using a relatively old 180-nm technology, the experimentally measured time delay and energy dissipation (per one pattern classification) are, respectively, 1 μs and 20 nJ [10], i.e., at least three orders of magnitude better than those reported for the best digital implementation of the same task, with a similar classification accuracy, using the 28-nm IBM's TrueNorth chip [11]. The maximum current per synapse in such design is still very high (three orders of magnitude above the noise floor of devices) indicating a significant room for improving the energy efficiency both at software and hardware levels, and enabling hardware implementation of larger deep learning models.

We have designed the largest (to the best of our knowledge) mixed-signal chip to run a convolutional neural network, with preliminary results provided in [7]. Floating-gate transistors are leveraged for both network weights storage, and performing input-weight multiplication operation in analog domain.

Computation on analog hardware offers speed and power efficiency at the expense of limited signal precision. Typical factors limiting the precision are thermal noise, quantization noise (for mixed signal circuits), device to device variation, circuit non-linearities, parasitic coupling, device programming accuracy, and programming retention. In addition, a small fraction of individual devices and/or connections is expected to fail (stuck at zero/one faults, etc.). Among these challenges, thermal noise is potentially the most significant for the ultra-low power analog circuits operating at room temperature, as it is the hardest to control. Even though neural networks "by design" are robust to reasonable amount of noise [15], it still degrades classification accuracy, and, therefore, an accurate simulation of hardware constraints during training must be performed to maximize performance of the model deployed on a chip.

The ideal method to make a neural network more robust to noise during inference is to train it in hardware [25-27]. This allows the model to adapt to hardware constraints and circuit nonidealities, while learning its weights using the standard gradient descent algorithm. However, adding the circuitry needed for backpropagation significantly complicates chip design and increases chip area, even though the training operation might be needed only once. In addition, implementing low power training procedure in analog hardware is a significant challenge, because signal precision needed to perform incremental weight updates is much greater than that needed for inference.

Chip-in-the-loop method [13, 28, 48] is sometimes used for inference only hardware as an approximation of training on chip. It involves transferring model weights to the chip, programming the chip, performing a forward computation pass on-chip, measuring the error, and then backpropagating this error in software and updating the weights on the conventional processor (CPU or GPU), before transferring them again to the chip. This process is slow and inefficient and is usually used only to fine tune the pretrained model for a small number of iterations.

In this work, we focus on making the network more robust to noise while training it in entirely in software. Fig. 1 demonstrates the main result, which we achieve through a combination of the accurate noise simulation and optimization of signal ranges: x-axis displays the maximum current for a synaptic device corresponding to a maximum weight. Not surprisingly, the crossbar power consumption is almost linearly proportional to the maximum weight current and so is the total

This work was supported in part by Semiconductor Research Corporation (SRC) and DARPA's UPSIDE program via BAE Systems, Inc

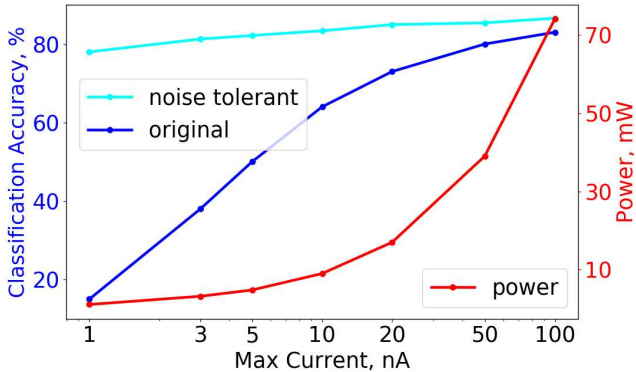

Fig. 1: Classification accuracy improvement (blue and cyan curves correspond to the first and last rows of Table II) and corresponding crossbar power consumption (red curve) for the network trained with and without noise tolerance enhancing methods.

power consumption since neurons are designed based on maximum pre-activation current. Lower maximum synaptic current leads to lower signal to noise ratio (SNR). Classification accuracy is measured on the test data in the presence of accurately simulated hardware noise. Cyan and blue curves show results for the model trained with and without methods described in Section V.

The novelty of this work and our specific contribution are:

- We have evaluated the impact of noise on the performance of a mixed-signal hardware convolutional neural network classifier, and explored the trade-off between classification accuracy and power consumption, reporting results on CIFAR-10 dataset. To the best of our knowledge, methods to deal with signal precision challenges specific to ultra-low power analog/mixed signal neural network hardware have never been evaluated.

- We developed an accurate noise model for our mixed-signal neural network hardware accelerator using mature embedded NOR flash memory technology. We believe our effort will serve as a useful case study for anyone designing ultra-low power hardware for neural network inference, regardless of their technology choice.

- We have evaluated several previously suggested methods to train the model to increase its robustness to noise, discovered a surprising benefit of batch normalization applied to outputs of the model, and developed a method to let the network itself learn the optimal signal ranges via gradient descent to increase noise robustness and improve signal to noise ratio.

## II. Related Work

Utilizing analog and mixed-signal hardware for running neural network based machine learning models has a long history [12, 14, 21-29]. Murray & Edwards in [29, 30] described one of the first attempts to simulate hardware specific noise during off-chip training for later deployment on a mixed signal chip. Same authors in [31, 32] proposed regularization of weight gradients to enhance noise tolerance of their model. A recent work in [33] provides a review of regularizer based methods to improve noise robustness, and proposes a new method, which we evaluate in Section V.

Hardware constraints other than noise have been evaluated in [34-37]. Some of such constraints for a convolutional neural network on CIFAR dataset have been evaluated recently in [45].

Circuit noise and other non-idealities in a mixed signal CNN accelerator have been considered and simulated in [48], which allowed the authors to anticipate the impact of such non-idealities on classification accuracy and choose circuit design parameters to minimize the impact, effectively "over-engineering" the hardware to make it immune to those non-idealities. In contrast, we adapt the software model, rather than hardware, to the anticipated circuit non-idealities. Two methods to enhance noise tolerance of this chip has been proposed in [49]: a form of chip-in-the-loop training, which is complementary to our work, and adding noise during training to the layer inputs. The authors show that it is more effective than adding noise to model weights. In our case, we add noise to layer preactivations, and our noise model is more sophisticated. In addition, we explore other training methods which further improve accuracy when combined with noise injection.

Similar to our goal, neural network quantization efforts aim to perform more efficient digital computation at the cost of reduced precision [39-41]. Especially relevant to our work are methods to optimize useful signal range of activations during training with quantization [42-44].

Recent study of neural network noise tolerance in [15] shows how training with one type of noise makes the model robust to other noise types. We test their methods and demonstrate that training with the same type of noise as what the model will experience during inference is still significantly more effective.

Excellent literature survey and overview of methods to improve neural network fault tolerance is provided in [50].

## III. Network Architecture

Our model is based on LeNet-5 architecture [16]. It uses 6 layers: Conv1 convolutional layer with 5x5 filters and 65 feature maps, Pool1 (max-pooling of 2x2 regions), Conv2 with 5x5 filters and 120 feature maps, Pool2 (max-pooling of 2x2 regions), FC1 (fully connected layer) with 390 neurons, and FC2 with 10 output neurons. Batch normalization [17] is supported in hardware and can be applied after each non-pooling layer. Activation function is ReLU. We test the accuracy of the model on CIFAR-10 dataset [20] of 32×32 pixel RGB images divided into 10 object classes. The dataset is split into 50k training images, and 10k test images, with the model trained on the training images, and tested on test images. We apply commonly used data augmentation methods during training: padding the input with two zero pixels, then cropping a random 32x32 region, and perform random horizontal flipping of images with probability of 0.5. No mean subtraction is performed (all input values are positive). The training is done off-chip in FP32 precision, using Nvidia Titan X GPUs, and the learned parameters achieving the highest test accuracy are then programmed into the mixed-signal chip.

For all experiments we use ADAM optimizer [18], batch size 64, the optimal initial learning rate varies depending on the experiment (determined via a grid search) and is scaled by 0.1 every 100 epochs, for 250 epochs total. Model parameters are initialized as suggested in [19]. Cost function is cross-entropy.

Each reported data point is an average of 5 training runs. The software model is implemented using Pytorch [46], and the code is available at http://github.com/michaelklachko/noisynet.

## IV. HARDWARE DESIGN

Conv1 layer is implemented with digital-input 4-bit merged-DAC (digital to analog converter) vector-by-matrix multipliers (VMMs). The choice of 4-bit input precision introduces quantization noise because in the original CIFAR-10 dataset image pixels are stored as three 8-bit integer values (one per RGB channel). To produce 4-bit dataset, we first divide each pixel value by 256, then we quantize the results to 4 bits. Table I shows that 4 bits per RGB channel is enough precision to achieve near optimal accuracy with this network.

TABLE I: CIFAR-10 IMAGE QUANTIZATION IMPACT

| Bits | 1 | 2 | 3 | 4 | 5 | 6 | 7 | 8 |
|---|---|---|---|---|---|---|---|---|
| Acc (%) | 74.7 | 83.6 | 86.8 | 88.1 | 88.2 | 88.3 | 88.2 | 88.2 |

Merged-DAC structure reduces the area/power overhead of data converters. In such topology, the current flowing in each synapse is given by

$$I_{syn,ij} = X_i(W_{ij}/W_{max})I_{s,max}$$

where $W_{max}$ is the largest absolute weight, obtained from software in the layer, $I_{syn,max}$ is the maximum designated synaptic current, and $X_i$ is the digital input. To facilitate differential weight mapping, we fully program (zero current upon maximum input voltage applied) either positive or negative merged-DAC devices. For embedded flash memories, our measurements based on 55 nm devices show that severe random telegraph noise is rarely observed (only one device among 140 devices). For high bandwidth applications, shot noise dominates the spectrum of current noise of devices. The spectrum density of current noise for a device conducting a subthreshold current $I_{syn,ij}$ is approximately $2qI_{syn,ij}$, where $q$ is electron charge. We assume $B_0$ = 250 MHz, a realistic equivalent noise bandwidth [7], to calculate the current noise of each device. The total current noise of all the devices in each VMM kernel (pre-activation) is the sum of all independent current noise sources and the variance of the noise added to each pre-activation in software is

$$\sigma_j^2 = (2qB_0)\left(\frac{W_{max}}{I_{s,max}}\right)\left(\sum X_i|W_{ij}|\right) \quad (1)$$

which is obtained by scaling the pre-activation referred current noise by $(X_iW_{ij})^2/(I_{syn,ij})^2$ to keep the dimensions meaningful and to make the SNR of each synapse equal in software and hardware. Note that the absolute factor stems from the fact that the current noise of all positive and negative synapses adds up in the output node. The network is simulated by adding a normally distributed noise to each pre-activation:

$$Y_{n,j} = Y_j + N(0, \sigma_j^2) \quad (2)$$

Other layers are implemented as analog-input analog-output current-mode VMMs [8]. Such topology employs current-mirror structure in which $I_i$ is an applied current to the gate-coupled array. To simplify the mapping, we can assume

$$X_i = I_i(X_{max}/I_{in,max}) \quad (3)$$

which indicate that currents are linearly mapped to the software values. For analog-input VMMs, $X_i$ is the $i^{th}$ input value which is always in $[0, X_{max}]$ due to the rectified linear activation function and $I_{in,max}$ is a designated maximum input current per layer. Given that, each synapse conducts $I_{syn,ij} = I_iW_{ij}$. Using the same procedure, we can find the equivalent pre-activation variance

$$\sigma_j^2 = (2qB_0)(X_{max}/I_{in,max})(\sum X_i(|W_{ij}| + |W_{ij}|^2)) \quad (4)$$

to use during model training. Note that the quadratic term is the impact of the peripheral floating-gate cell in current mirror topology. See [8] for more information.

## V. TRAINING THE MODEL

The main factors affecting the classification accuracy of our model are: device noise, quantization noise, weight programming precision, and non-linearities of the devices/circuit. Ex-situ trained mixed signal neural networks are less prone to device-to-device variations because each device (weight) is programmed individually, which also allows to compensate variations and resolve mismatch issues in circuits. Impact of circuit/device non-linearities will be explored in future work.

Eqs. 1 and 4 specify the amount of device noise for the first layer and the remaining layers, respectively. Given a particular power budget, we assign maximum current values, i.e. $I_{s,max}$ for the first layer, and $I_{in,max}$ for the remaining layers. (Note that each layer can have a different $I_{max}$ value.) These values determine the thermal noise floor and scale the amount of noise that will be added to the pre-activations in each layer during inference (per Eq. 2). Note that noise magnitude is also dependent on signal ranges (weights range for the first layer) and has a non-linear relationship with model parameters. Therefore, we have two objectives:

1. Improve the noise tolerance of the model using various regularization methods.

2. Improve the signal to noise ratio, i.e. for a given $I_{max}$, as model weights change during training, we want to reduce $\sigma_j^2$ relative to $Y_j$, on average.

We start with establishing a baseline for the classification accuracy in the absence of any hardware constraints. Using hyperparameters described in Section III, with learning rate 0.0005, batch normalization in Conv1, Conv2, and FC1 layers, weight decay (L2) penalty of 0.0005 in all layers, and dropout 0.1 after Conv2 and FC1 layers we achieve 88.1% accuracy on the test set. If we remove weight decay and dropout (used to reduce overfitting) the baseline accuracy drops to 86.5%.

In all experiments that follow, we will be evaluating classification test accuracy when we inject noise according to Eqs. 1 and 4, referred to as "accurate noise", during inference. We will use the same value of $I_{max}$ in all the layers. In Section

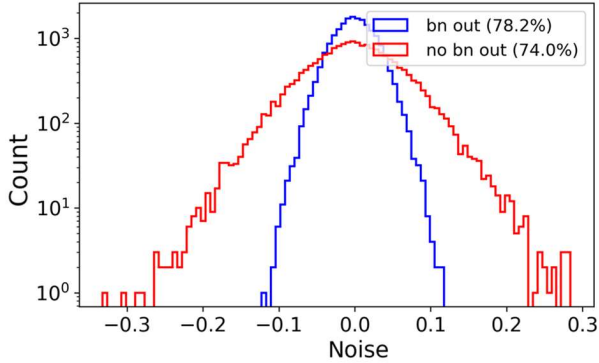

Fig. 2: Effect of batch normalization applied to model outputs on SNR: histograms of signal distortions divided by signal range in the last layer $n_{ij} = (Y_{n,j} - Y_j)/(Y_{max} - Y_{min})$ with and without batchnorm, and the corresponding test accuracies.

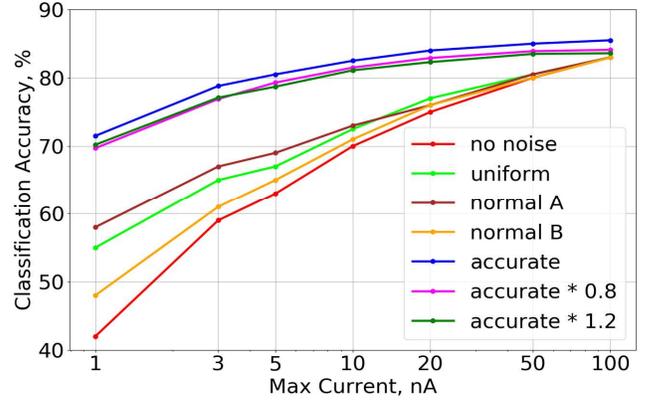

Fig. 3: Comparison of different noise types injected during training, with accurately modeled noise injected during testing. Red curve corresponds to "Large LR + BN outputs" row in Table II.

VI we will show how assigning different $I_{max}$ in each layer can lead to better power efficiency.

Batch normalization is applied after the noise has been added to a pre-activation. If an experiment results in different signal distributions at train and test times (e.g. no noise during train, and noise during test), batch statistics are calculated separately during test time.

### A. Weight Regularization

If we use the baseline model while injecting accurate noise during inference, we achieve accuracy corresponding to blue curve in Fig. 1. Increasing learning rate to 0.02 significantly improves the accuracy, possibly because large learning rates are more likely to find flat regions of the loss function landscape, making weights more robust to perturbation. As the amount of noise decreases, this becomes a disadvantage, as the model is unable to find more fine-grained/narrow minima regions because of the large update steps.

Next, we tried adding batch normalization after FC2 layer (model outputs, right before softmax), which turns out to be surprisingly effective. As far as we know, batch normalization has never been used in the output layer of a classifier, probably because it does not appear to have any beneficial effect in a noise-free environment. Batchnorm in the output layer might allow higher learning rates to be more effective in finding flat regions of the loss landscape. Moreover, it results in an SNR improvement as shown in Fig. 2. In all following experiments, unless specified otherwise, we will use batch normalization in all layers (including the output layer).

Several regularization methods to improve noise tolerance have been evaluated in [33]: penalizing weight growth ($E_1$), weight gradient growth ($E_2$), and weight second order gradient growth ($E_3$). We have tested each one in our model, except for $E_3$, because calculating diagonals of Hessians during training for a model with 1.3M parameters is impractically slow (the results in [33] are reported for a very small MLP network). We have tried a simplified method $E_3'$ instead, where we calculate the sum of gradients for all weights, then penalizing growth of derivatives of that sum in respect to each weight:

$$E_3' = \sum (d \sum (\frac{dC}{dW_i})/dW_i)^2$$

However, no significant improvement has been observed when applying this penalty alone or in combination with other methods. As shown in Table II, $E_1$ and $E_2$ are quite effective when combined with large learning rate and batch normalization in the last layer. We have also tested a variant of $E_2$ applied to pre-activation gradients, rather than weight gradients, and obtained similar results. $E_1$ is only effective when applied to the first layer weights, we believe this is mostly due to an improvement in SNR (note how in Eq. 1 decreasing $W$ also decreases $W_{max}$, leading to faster decrease of $\sigma_j^2$ than in Eq. 4). We also noticed that aggressive weight gradient clipping helps when used in combination with $E_1$. The effect of gradient clipping is intriguing and requires further investigation. As expected, $E_2$ makes the model more robust to noise, however combining $E_1$ and $E_2$ does not lead to further accuracy improvement. This is contrary to what has been reported in [33] and might be caused by $E_1$ trying to decrease weights while $E_2$ is trying to increase them in the same update step, thus cancelling benefits of each. Similar training dynamic was observed in [51]. Further investigation is required to determine if it's possible to use gradient regularizers while preventing weight growth.

### B. Importance of Accurate Noise Model

It's been suggested in [15] that networks trained with one type of noise are robust to other types of noise. We tested noise sampled from uniform and normal distributions, with pre-activation magnitude dependent variance as well as signal range dependent variance. In all cases we use accurate noise model at test time. Note that in [15] noise was added to weights, while we are adding noise to pre-activations. Distorting weights during training while distorting pre-activations at test time led to slightly inferior results in our experiments for all noise types. Fig. 3 shows results for the following types of noise (added to each pre-activation during training, before batch normalization):

a. Noise sampled from uniform distribution with range proportional only to the dynamic range of pre-activation values in each layer.

b. Noise sampled from normal distribution (normal A) with variance proportional only to the dynamic range of all pre-activation values in each layer.

c. Noise sampled from normal distribution (normal B) with variance for each pre-activation proportional to the pre-activation magnitude.
  d. Noise according to our noise model (accurate noise).
  e. Noise according to our model, with variances in Eq. 1 and Eq. 4 increased and decreased by 20%.

Again, batch normalization has been also applied to the outputs of the model, and batch statistics were calculated separately during train and test times, because of a potential distribution mismatch. Optimal scaling factors for each noise type were found through sweep search, independently for every experiment. Note that in the absence of an accurate noise model, there's no way to evaluate the effectiveness of training with these noise types, other than testing it directly on the target hardware. It's clear from the figure that accurate noise modeling during training is important as it significantly outperforms other noise types, and even slight deviations from the accurate noise model lead to accuracy degradation. Surprisingly, we found that training with pre-activation magnitude dependent uniform noise (referred to as "StochM " in [15], not shown in the figure), implemented as multiplication of each pre-activation value $y$ by a uniform random variable in the range $[0.5y, 2y]$, together with dropout to randomly set a fraction of pre-activations to zero, never produced results better than training with no noise at all.

We have also tried binarizing weights, as performed in [15] and [47], to evaluate noise tolerance of a binary weight network. However, even without injecting any noise, baseline accuracy of our model dropped from 88.1% to 81.4%. This might seem surprising, given that the authors of [39] claim no accuracy drop on CIFAR-10, and we indeed verified it by running their code [47] and seeing no accuracy drop on CIFAR-10 when using their models. The reason is difference in model sizes: Alexnet model in [47] has 44M parameters, and the network described in [15] has 13M), whereas our network has only 1.3M parameters, so we have the effect reported in [5]: the larger (and possibly more redundant) the model, the less susceptible it is to quantization noise. We tried both binary weights and binary activations networks and the accuracy was worse than 'no noise' baseline in Fig. 3 for all values of current.

## C. Weight and Actvation Clipping

Eqs. 1 and 4 suggest that limiting largest weights and largest activations, respectively, might lead to improved SNR in the corresponding layers. Fig. 4 shows how clipping of the first layer weights during training improves the accuracy for various clipping thresholds (threshold $t$ defines the range $[-t, t]$ of allowed weight values). Each curve shows the difference between classification accuracy for a given clipping threshold and the accuracy when no clipping is done (rightmost point). When the clipping is too aggressive, the narrow range of weight values makes it more difficult to find optimal ones. More importantly, weight programming precision is limited, as explained in Section 5E. On the other hand, allowing the full range of weight values in the first layer leads to poor SNR, as can be seen in Fig. 5, where we plot relative errors: distortions of each pre-activation divided by a range of pre-activations for the first layer. We estimate the range as a difference between 99$^{th}$ and 1$^{st}$ percentiles, because the full range fluctuates too much between training runs.

Effects of activation clipping (Fig. 6) can be explained in a similar manner: too aggressive clipping requires increased weight precision, which is limited by weight programming precision, while less clipping, again, degrades SNR (as shown in Fig. 7). In addition, clipping noisy signals, as long as not too many of them are constrained, might have a beneficial regularization effect similar to batch normalization.

Table II provides comparative results for all the methods we tried. Injecting accurate noise during training is the most effective method, followed by clipping, which is more effective than the gradient growth regularizer ($E_2$). Combining $E_2$ with clipping leads to significant improvement over each one of them in isolation, however this holds only for the noisiest scenario ($I_{max}$ = 1 nA), and for the remaining values of $I_{max}$, $E_2$ leads to degradation of accuracy (compared to clipping only). Combination of clipping and noise further improves the accuracy. Unfortunately, combining of all three of these methods (clipping, noise, and $E_2$) does not improve the accuracy further. As has been shown in [38] injecting noise might have a similar effect to L2 regularization ($E_1$). Therefore, we might have the same opposing effects trying to combine $E_2$ and noise as what we observed when combining $E_1$ and $E_2$. Another reason might be that adding $E_2$ when training with noise leads to too much regularization, as evidenced by degradation of training accuracy. Finally, batch normalizing model outputs improves the accuracy further for all combinations and is especially effective for low current values. Note that the results in the last row of Table II for $I_{\max} = 100$ nA are still significantly below the baseline (88.1%) – this is explained by the lack of regularization that we used for the baseline to prevent overfitting. Indeed, adding the same regularization (L2 = 0.0005 and dropout 0.1) improves the accuracy to 87.94%.

TABLE II: EFFECT OF VARIOUS TRAINING METHODS ON NOISE TOLERANCE

| Maximum Input Current, nA | 1 | 3 | 5 | 10 | 20 | 50 | 100 |
|---|---|---|---|---|---|---|---|
| Baseline | 14.93 | 38.21 | 50.02 | 64.23 | 73.05 | 80.14 | 82.94 |
| Large LR | 38.56 | 55.33 | 60.28 | 67.55 | 74.59 | 80.98 | 83.45 |
| Large LR + BN outputs | 42.78 | 59.63 | 63.33 | 70.59 | 75.85 | 81.34 | 83.61 |
| BN outputs + E1 | 47.50 | 63.28 | 70.56 | 75.23 | 79.47 | 82.61 | 84.11 |
| BN outputs + E2 | 53.52 | 63.59 | 68.44 | 74.05 | 78.36 | 81.43 | 83.20 |
| Clipping | 58.35 | 73.95 | 77.41 | 80.81 | 82.47 | 84.48 | 85.13 |
| BN outputs + clipping | 62.83 | 74.43 | 77.70 | 80.40 | 82.54 | 84.03 | 84.19 |
| Noise | 65.31 | 75.61 | 78.85 | 81.85 | 83.34 | 84.75 | 85.52 |
| BN outputs + E2 + clipping | 67.28 | 74.43 | 77.70 | 80.40 | 82.54 | 84.03 | 84.19 |
| Noise + BN outputs | 71.75 | 78.39 | 80.23 | 82.62 | 84.12 | 84.84 | 85.45 |
| Noise + clipping | 73.50 | 80.14 | 81.74 | 83.07 | 84.63 | 85.35 | 86.36 |
| Noise + BN outputs + clipping | 78.00 | 81.27 | 82.18 | 83.37 | 84.95 | 85.50 | 86.58 |

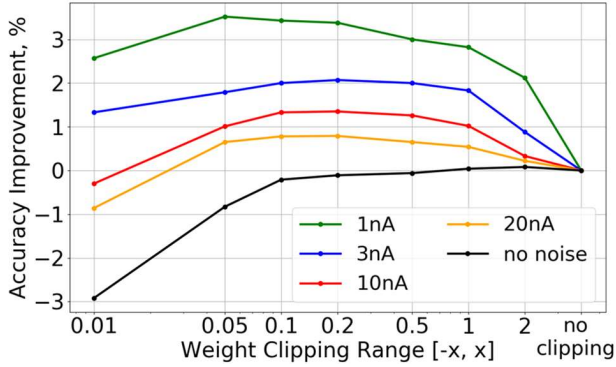

Fig. 4: First layer weight clipping (absolute accuracy improvements relative to no clipping).

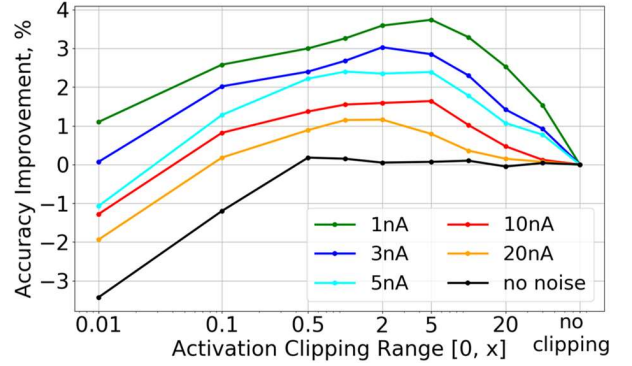

Fig. 6: First layer activations clipping (absolute accuracy improvements relative no clipping).

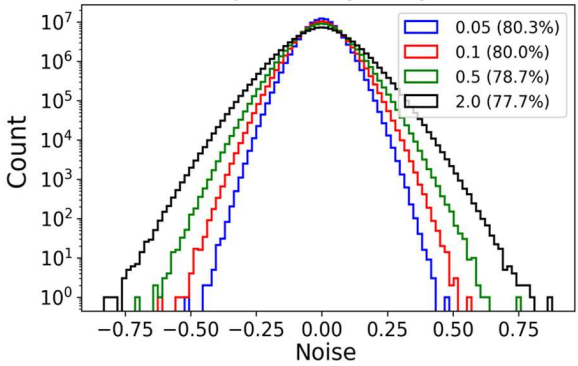

Fig. 5: Signal distortions for different first layer weight clipping thresholds: datapoints are distortions applied to individual first layer pre-activations, divided by the layer signal range. For example, red curve shows distribution of signal distortions when weights are clipped in [-0.1, 0.1] range, normalized by the corresponding signal range, and with the corresponding classification accuracy of 80%. $I_{\max} = 1$ nA (noise is injected in the first layer only).

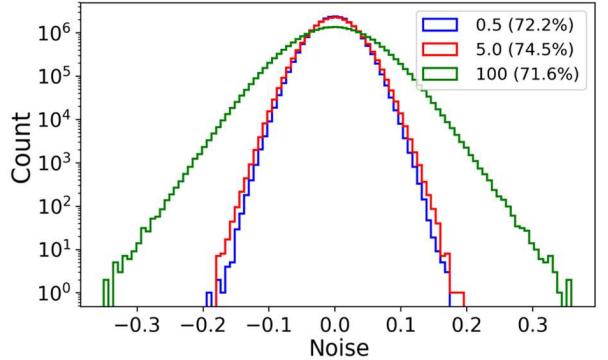

Fig. 7: Signal distortions improvement for different second layer input clipping: datapoints are distortions of each second layer pre-activation divided by signal range. $I_{\max} = 1$ nA in all layers.

### D. Learning Optimal Clipping Thresholds

The optimal clipping thresholds can be different for each layer. Determining these thresholds manually for each layer is a very time-consuming task, even for just three layers, because, for example, an optimal threshold $Y_{\max}$ for Conv1 found while not clipping inputs in Conv2 and FC1 is not necessarily the optimal value when we perform clipping in all layers. To find the best thresholds in all layers, it was necessary to first find optimal values for all layers in isolation, then adjusting each one again, because when we change signal range in one layer distributions of the signals (and noise) in other layers change. This iterative clipping thresholds adjustment process would be unfeasible for deeper models. Moreover, optimal thresholds might depend on the choise of methods used during training (e.g. $E_1$ or $E_2$). Therefore, we want the network to jointly learn the optimal clipping thresholds for all layers during training.

One way to achieve this is to accumulate gradients for the activations being clipped by threshold $Y_{\text{thr}}$, for each batch of training images, and let this sum of gradients guide the amount of change for the threshold itself:

$$\frac{dC}{dY_{\text{thr}}} = \Sigma \frac{dC}{dY_i}, \quad for\ Y_i > Y_{\text{thr}}$$

where $C$ is the cost function. Learning the optimal thresholds in this manner provide information about how sensitive each layer is to clipping. Next, we introduce a single tuning parameter for all layers, which determines the amount of force to apply to push all the thresholds down. This is accomplished with L2 penalty added to the loss function:

$$C = C + \alpha \sum (Y^l_{thr}/I^l_{max})^2$$

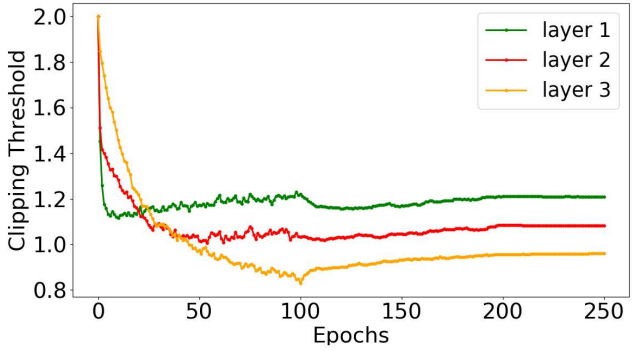

Fig. 8: Optimal activation clipping thresholds learned during training, $I_{\max} = 1$ nA in all layers, $\alpha = 0.01$.

where $\alpha$ is the strength of the penalty. Note that each term is scaled by the max current in that layer, because each layer can have a different $I_{\max}$, and the optimal clipping region depends on $I_{\max}$ (see Fig. 6).

Now instead of searching for optimal clipping bounds, one layer at a time, we can just search for the optimal $\alpha$ parameter. Fig. 8 shows an example of the activation thresholds evolution during training (all thresholds were initialized to higher than desired value). Most importantly, this method results in equal or better accuracy than what we get with manually optimized thresholds one layer at a time.

### E. Weight Programming Precision

Experimentally, we have observed that by integrating enough in the read-out circuitry, we can program the device within $I_{res} = 0.1$ nA resolution in deep subthreshold region. To accurately model this in software, we add a randomly sampled $\Delta W$ from a uniform distribution to the nominal weight value $W_i^n$. Hence, the weight value in the software is calculated by

$$W_i = W_i^n + U\big(W_i^n - I_{res}/I_{in,\max},\, W_i^n + I_{res}/I_{in,\max}\big)$$

Note that the maximum synaptic current determines the resolution of weight programming. For example, if $I_{in,\max} = 3$ nA, the resolution of weight values in software is 0.033.

We found that this effect should not be simulated during training when we already inject significant amount of pre-activation noise ($I_{\max} < 10$ nA), because it seems to reduce the learning capacity of the model. Instead, we simulate it only during inference. Results in Table III show the absolute drop in classification accuracy (difference between testing with and without random weight distortions). Clearly, this impact should be taken into an account when choosing (or learning) the optimal clipping threshold for the first layer weights, strength of weight decay ($E_1$) penalty, and/or learning rate.

## VI. Power Optimization

After noise-optimal network parameters are found, given target accuracy and power budget, current values for each layer should be adjusted so that both requirements are satisfied. We start with noise sensitivity test for each layer, where we inject noise ($I_{\max} = 10$ nA) into a baseline model one layer at a time and evaluate its impact on the accuracy. Table II shows that the first layer is by far the most sensitive to distortions, and the second layer consumes the most power. Given this information, we can adjust $I_{\max}$ for the first three layers in proportion to their importance. The output layer consumes very little power, so we can increase $I^4_{max}$ to maximize SNR. Setting $I^1_{max} = 1.8$ nA, $I^2_{max} = 1.4$ nA, $I^3_{max} = 5$ nA, $I^4_{max} = 40$ nA, and using aggressive weight decay penalty in the first two layers, we achieved 80.3% accuracy with just 1.4 mW power budget, and 87.1% with 6 mW. Ideally, we would like to let the network learn the optimal balance of noise distribution between layers given a specific power budget.

TABLE IV: Layer Noise Sensitivity and Power Consumption

| Layer | Conv1 | Conv2 | FC1 | FC2 |
|---|---|---|---|---|
| Noise sensitivity | 11.3% | 4.6% | 3.8% | 1.2% |
| Power consumption | 35% | 59% | 5.6% | 0.4% |

## VII. Conclusion

Attempting to run neural networks on ultra-low power mixed signal hardware introduces significant amount of signal distortions, which degrade classification accuracy. We have shown that combining accurate simulation of the distortions during off-chip training of the model, adaptive signal range optimization, and normalizing the outputs of the model makes the network significantly more robust to noise, improves SNR, and ultimately allows better control over the power-accuracy tradeoff. In future works, we plan to study other circuit/device non-linearities and explore methods to train a network so that it can learn the optimal parameters given a power budget or target accuracy.

TABLE III: Weight Programming Precision Impact

| Maximum Weight | 0.05 | 0.1 | 0.3 | 0.5 | 1 |
|---|---|---|---|---|---|
| $I_{\max} = 1$ | 27% | 19% | 7.3% | 1.5% | 0.6% |
| $I_{\max} = 3$ | 12% | 5.7% | 2.8% | 0.6% | 0.2% |
| $I_{\max} = 5$ | 3.6% | 1.1% | 0.7% | 0.1% | 0.1% |